\documentstyle[twoside,fleqn,espcrc2,epsf]{article}

\newcommand{\AmS}{{\protect\the\textfont2
  A\kern-.1667em\lower.5ex\hbox{M}\kern-.125emS}}
\def\d{\partial}

\def\a#1#2{a^{\rm #1}_{#2}}

\def\BE{\begin{equation}}
\def\EE{\end{equation}}
\def\BEA{\begin{eqnarray}}
\def\EEA{\nonumber\end{eqnarray}}

\title{Hodge gauge fixing in three dimensions}

\author{J. E. Hetrick
\address{Physics Department, University of Arizona, Tucson, AZ 85721}}
       
\begin{document}

\begin{abstract}
A progress report on experiences with a gauge fixing method proposed
in LATTICE 94 is presented. In this algorithm, an SU(N) operator is
diagonalized at each site, followed by gauge fixing the diagonal
(Cartan) part of the links to Coulomb gauge using the residual abelian
freedom. The Cartan sector of the link field is separated
into the physical gauge field $\alpha^{(f)}_\mu$
responsible for producing $f^{\rm Cartan}_{\mu\nu}$, the pure gauge
part, lattice artifacts, and zero modes. The gauge transformation to
the physical gauge field $\alpha^{(f)}_\mu$ is then constructed and
performed. Compactness of the fields entails issues related to
monopoles and zero modes which are addressed.
\rightline{AZPH-TH/96-18}
\end{abstract}
\maketitle

\section{The Method}
While gauge fixing is a central tool of lattice simulations, the
effect of lattice artifact ``Gribov copies'' remains a delicate issue,
particularly in chiral fermion models which often rely on gauge fixing
essentially.

In \cite{dFH} we proposed a method of gauge fixing to an t'Hooft like
gauge which was called {\it Gauge fixing by Hodge decomposition}. We
work here on a spatial torus in 3-dimensions, thus this presentation
is for Coulomb gauge (the algorithm is done in parallel on each time
slice). Landau gauge generalizes similarly. The method is as follows.
\begin{itemize}
\item Diagonalize some operator which transforms adjointly,
${\cal O}_x^\prime = G_x^\dagger {\cal O}_x G_x$, at each site.
(The operator used here is the spatial sum of plaquette 
clovers and their adjoints at each site.)
\item Define an Abelian $\alpha_\mu$ field from the links.
\item Decompose $\alpha_\mu(x) = \alpha^{(f)}_\mu + \d_\mu\phi +
H_\mu + w_\mu$, and solve for the physical field
$\alpha^{(f)}_\mu$ which minimally produces the plaquettes 
and open (monopole) strings $f_{\mu\nu}$.
\item Add the continuum zero-mode: $\alpha^{(f)}_\mu + w_\mu$.
\item Construct and perform the $U(1)$ gauge transformation which moves
the links from $\alpha_\mu$ to $\alpha^{(f)}_\mu + w_\mu$.
\end{itemize}

\subsection{Diagonalization of ${\cal O}$}
An iterative method can be used in which the operator ${\cal O}$ is
hit with successive $SU(2)$ subgroup gauge transformations
that each minimize the modulus of resulting off-diagonal terms. If
\BE
{\cal O}^\prime =
G^\dagger
\left(\matrix{a_{11} & a_{12}\cr
              a_{21} & a_{22}\cr}\right)
G
\EE
then $\theta = \tan^{-1}(-2A/B)/4$~~
minimizes $|a^\prime_{12}| + |a^\prime_{21}|$. In the two cases:

\leftline{\bf $\sigma_x$ case:}
\BE
G = \left(\matrix{\cos\theta &  i \sin\theta\cr
             i\sin\theta & \cos\theta\cr}\right)
\EE
\BEA
A &=& \a{re}{11}\a{im}{12} - \a{im}{11}\a{re}{12} -
\a{re}{11}\a{im}{21} + \a{im}{11}\a{re}{21}\cr
   &&+ \a{re}{12}\a{im}{22} - \a{re}{21}\a{im}{22} -
\a{im}{12}\a{re}{22} + \a{im}{21}\a{re}{22}\cr
B &=& (\a{re}{11})^2 + (\a{im}{11})^2 + (\a{re}{22})^2 + (\a{im}{22})^2\cr
&&-(\a{re}{12})^2 + (\a{im}{12})^2 - (\a{re}{21})^2 - (\a{im}{21})^2\cr
&& +~2\a{re}{12}\a{re}{21} + 2\a{im}{12}\a{im}{21}
- 2\a{re}{11}\a{re}{22} - 2\a{im}{11}\a{im}{22}
\EEA

\leftline{\bf $\sigma_y$ case:}
\BE
 G = \left(\matrix{\cos\theta &  -\sin\theta\cr
             \sin\theta & \cos\theta\cr}\right)
\EE
\BEA
A &=& \a{re}{11}\a{re}{12} + \a{im}{11}\a{im}{12} + 
\a{re}{11}\a{re}{21} + \a{im}{11}\a{im}{21}\cr
 &&  - \a{re}{12}\a{re}{22} - \a{im}{12}\a{im}{22} 
- \a{re}{21}\a{re}{22} - \a{im}{21}\a{im}{22}\cr
B &=& (\a{re}{11})^2 + (\a{im}{11})^2 + (\a{re}{22})^2 + (\a{im}{22})^2\cr
&&-(\a{re}{12})^2 + (\a{im}{12})^2 - (\a{re}{21})^2 - (\a{im}{21})^2\cr
&& -~2\a{re}{12}\a{re}{21} - 2\a{im}{12}\a{im}{21}
- 2\a{re}{11}\a{re}{22} - 2\a{im}{11}\a{im}{22}
\EEA
For ${\cal O} \in SU(2)$ or $su(2)$, only one of each gauge
transformation is required. For $SU(3)$ operators, about 10
iterations are required to get $\sum_{\rm off-diagonal}|a_{ij}| < 10^{-7}$.

\subsection{The Residual Abelian Fields}

We must define residual $U(1)$ fields, which represent the part of
$A^a_\mu$ in the Cartan subgroup in the continuum limit.
For $SU(2)$ we use:
\BE
\alpha_\mu = \tan^{-1}(e_3/e_0)
\EE
where $~U_\mu = e_0 + i e_k \sigma_k$.

\medskip
For SU(3) where a link has diagonal elements
$\{e^{i\beta^1},e^{i\beta^2},e^{i\beta^3}\}$,
%
the angles
\BEA
\alpha^1_\mu &=& (2\beta^1 - \beta^2 - \beta^3)/3,
~~~~{\rm and}\cr
\alpha^2_\mu &=& (2\beta^2 - \beta^1 - \beta^3)/3
\EEA
are suitable Cartan fields.

\section{Hodge Decomposition of $\alpha_\mu$}

Any lattice vector field $\alpha_\mu(x)$ can be
uniquely decomposed into:
\BE
\alpha_\mu(x) = \alpha^{(f)}_\mu + \d_\mu\phi + H_\mu + w_\mu
\EE
where $\d_\mu$ ($\d^\dagger_\mu$) is the lattice forward (backward)
derivative
\begin{itemize}
\item $\alpha^{(f)}_\mu$ is the physical part of the field
and soley responsible for producing\\ 
$f_{\mu\nu} = \d_\mu \alpha^{(f)}_\nu - \d_\nu \alpha^{(f)}_\mu$.
Conveniently, $\alpha^{(f)}_\mu$ also naturally satisfies the Landau gauge 
condition:~~$\d_\mu \alpha^{(f)}_\mu = 0$.

\item $\d_\mu \phi$ is the pure gauge part.
\item $H_\mu$ is the lattice harmonic part responsible for Dirac string loops
(and is the major source of Gribov copies).
\item $w_\mu$ is the continuum harmonic part for a torus, ie. a constant. 
\end{itemize}

The decomposition is formal at this stage, but we remark that much of
the work in solving for the lattice $\alpha^{(f)}_\mu$ is in
identifying the neccessary parts of $H_\mu$ which must be kept.

\subsection{Solving for $\alpha^{(f)}_\mu$}

$\alpha^{(f)}_\mu$ is determined (up to zero-modes) by
\BE
\label{eq1}
\alpha^{(f)}_\mu = \Delta^{-1} \d_\nu^\dagger f_{\nu\mu},
\EE
or
\BE
\widehat\alpha^{(f)}_\mu(k) = \frac{ik_\nu^\dagger}{-k^2} 
\widehat f_{\nu\mu}(k),
\EE
in Fourier space. $k_\mu$ is the lattice momentum: $2\sin(\pi a
n_\mu/L_\mu)$. It is thus very easy to find $\alpha^{(f)}_\mu$ from
$f_{\mu\nu}$~ by FFT; however we must build the correct $f_{\mu\nu}$
in stages.

Since we want $\alpha^{(f)}_\mu$ to minimally reproduce the plaquette angles
$f^{\rm plq}_{\mu\nu}$: 

\medskip
$\bullet$~~~~~We first set $f_{\mu\nu} = f^{\rm plq}_{\mu\nu}$.

\section{Monopoles}
For compact gauge fields, $\alpha^{(f)}_\mu$ must also the reproduce
gauge invariant monopoles which are the ends of open lengths of Dirac
string; thus we need to find all monopole sites on the dual
lattice. Since the Dirac strings connecting these monopoles are gauge
variant, we define a monopole-anti$\cdot$monopole pairing which
minimizes the string length between pairs. This can be done by
simulated annealing for instance, but since this pairing is not
unique, this step is a source of Gribov ambiguity. We could also
connect pairs in order of finding them (violating rotational
invariance), which is however fast and unique. Then,

\medskip
$\bullet$~~~We next add to $f_{\mu\nu} = f^{\rm plq}_{\mu\nu} + 
f^{\rm mnple}_{\mu\nu}$
\medskip

in otherwords, for each plaquette pierced by a Dirac string (as given
by our minimal monopole pairing), we add $\pm 2\pi$ to $f_{\mu\nu}$. 

\subsection{Zero-modes of $f_{\mu\nu}$: ~~globally wrapping string}

Due to compactness again, $f_{\mu\nu}$ may have a zero-mode. This
zero-mode can be viewed as a length of non-contractable Dirac string,
ie. one that stretches across the torus. 
$$
\frac{1}{2\pi L^3}\int dx^\rho dx^\mu dx^\nu (f^{\rm plq}_{\mu\nu} + 
f^{\rm mnple}_{\mu\nu}) \equiv c^\rho \not= 0
$$
($\mu,\nu \perp \rho$) indicates the occurance of $c^\rho$
global non-contractable Dirac strings in the $\rho$-direction
in the initial configuration
which must be added to $f_{\mu\nu}$.

We are at liberty to place these strings wherever we like, either randomly
for pseudo translational invariance, or along some particular
axis so that we know where they are. Along these strings we add $\pm
2\pi$ to the $f_{\mu\nu}$ perpendicular to the path
so that $f_{\mu\nu} = f^{\rm plq}_{\mu\nu} + 
f^{\rm mnple}_{\mu\nu} + f^{\rm global}_{\mu\nu}$

We are now ready to solve eq. \ref{eq1}, which will give the minimal
$\alpha^{(f)}_\mu$ that reproduces the plaquette angles, monopole,
and global strings, and which is in Landau gauge:
$$
\framebox[7cm]{$
\alpha^{(f)}_\mu = \Delta^{-1} \d_\nu^\dagger 
(f^{\rm plq}_{\nu\mu} + f^{\rm mnple}_{\nu\mu} + f^{\rm global}_{\nu\mu}).
$}
$$

\subsection{One more zero-mode $w_\mu$}

There is one more zero mode that is undetermined by eq. \ref{eq1}
which is responsible for producing the correct Wilson loops. 
After solving for $\alpha^{(f)}_\mu$ we must add to it the correct
constant $w_\mu$
\BE
w_\mu = \frac{1}{L^3} \int\int dx^\rho dx^\nu\int
dx^\mu  \left(\alpha_\mu - \alpha^{(f)}_\mu\right)
\EE
in order that Wilson loops are preserved mod($2\pi$).

This is the zero-mode of the original $\alpha_\mu$ field, and does not
contribute to $f_{\mu\nu}$.

\section{The Gauge Transformation}

Because of the zero-modes, the only way to find the gauge
transformation taking us from $\alpha_\mu \longrightarrow
\alpha^{(f)}_\mu + w_\mu$ is by constructing a tree, or in otherwords
integrating the equation
$$
\d_\mu g = (\alpha^{(f)}_\mu + w_\mu) - \alpha_\mu 
$$

\section{Tests}

Figure 2 shows three extremization gauge fixed copies derived from
the initial configuration in figure 1. In figure 3, the {\em only} two
copies obtained by the Hodge method are displayed. The $SU(2)$
starting configuration is relatively smooth though, generated at $\beta =
44$, followed by a random gauge transformations.

\begin{figure}[htb]
\epsfxsize=6cm
\epsffile[83 255 363 508]{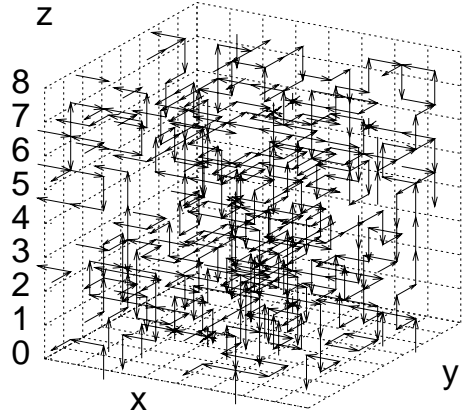}
\label{fig:1}
\caption{Initial gauge field configuration and string topology, on
slice $t=0$. The ``sum of clovers'' of an $SU(2)$ field was first
diagonalized according to the introduction, then a random gauge
transformation was applied.}
\end{figure}

\begin{figure}[htb]
\epsfxsize= 6cm
\epsffile[70 83 277 666]{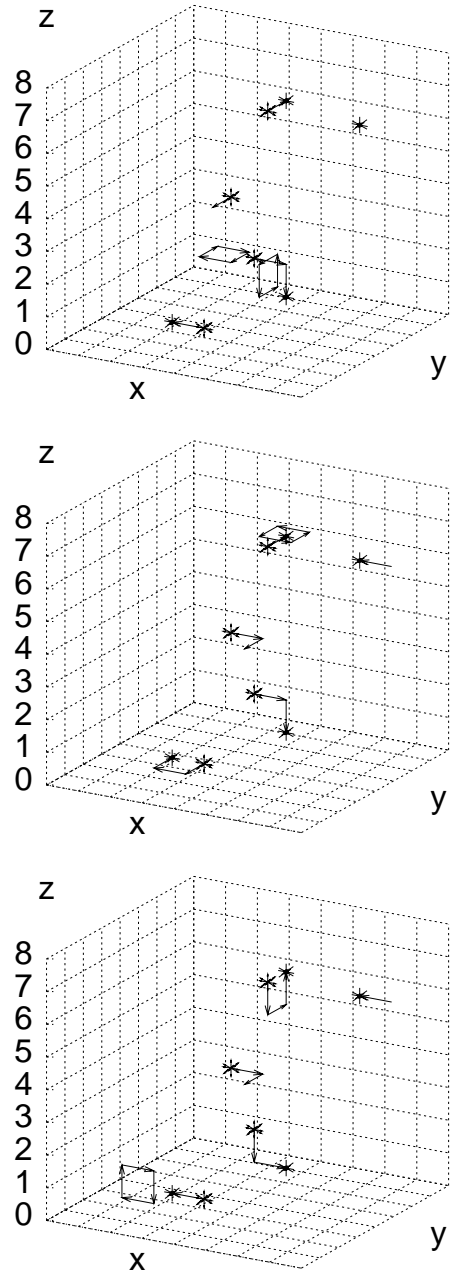}
\caption{
Three of the many Gribov copies from gauge fixing by extremization.}
\label{fig:2}
\end{figure}

\begin{figure}[htb]
\epsfxsize= 6cm
\epsffile[70 86 280 465]{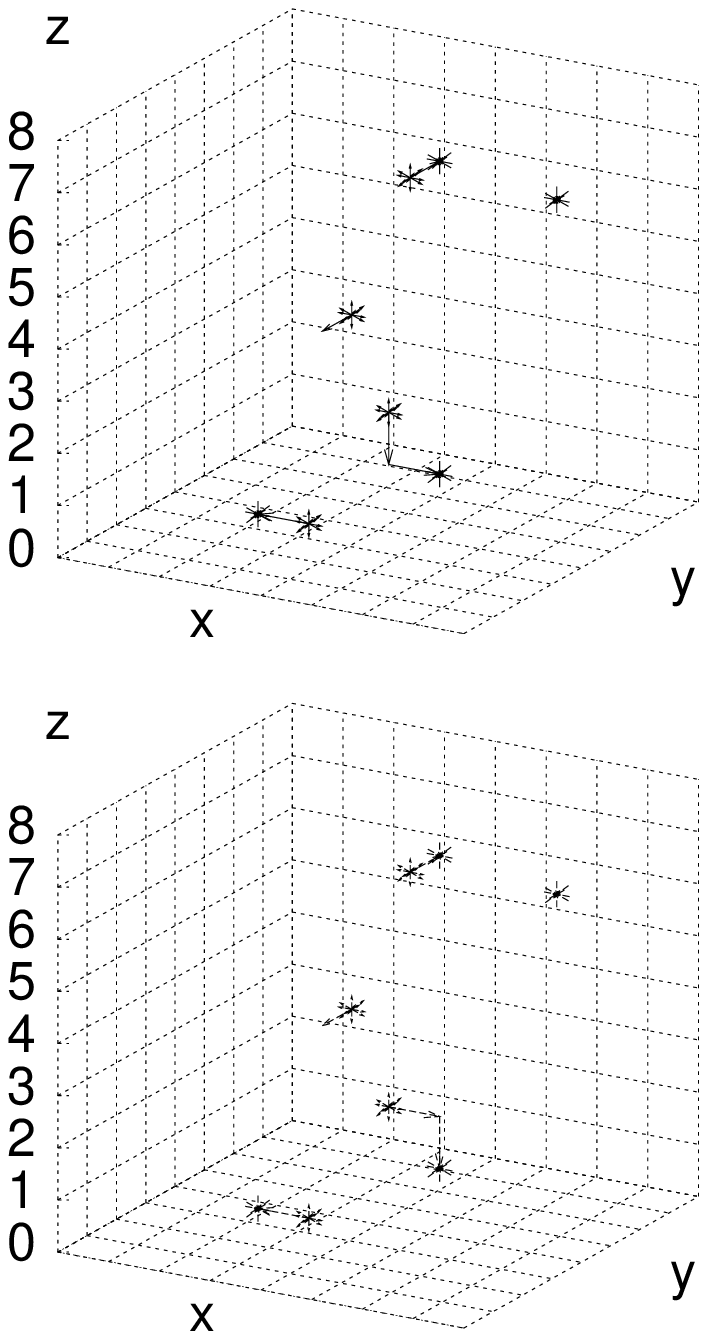}
\caption{Hodge method: only 2 final configurations occur.
\vskip 6cm}
\label{fig:3}
\end{figure}

\section{Conclusions}

\begin{itemize}

\item The method seems to work resonably well; it returns a
uniquely gauge fixed configuration, up to the connectivity of monopole
pairs. However, the Cartan sector of $SU(N)$ fields at typical $\beta$
values is fairly rough, and thus the monopole density is
also relatively high.

\item Closed (contractable) loops of Dirac string are removed, 
which are a primary  source of Gribov copies in extremization methods.

\item For high monopole densities, simulated annealing seems to give poorer
connectivity for the large number of monopole pairs than
extremization, ie. extremization uses less string to connect
monopoles. 

\item For rough fields, $|\alpha^{(f)}_\mu|$ can occasionally be larger
than $\pi$. This means that the physical (non-compact) field
$\alpha^{(f)}_\mu$ is ``clipped'' by compactness, resulting in a site
with $\d_\mu \alpha_\mu = \pm 2\pi$.
\end{itemize}

\end{document}